# Deep learning of spectra: Predicting the dielectric function of semiconductors

Malte Grunert ,[*,†] Max Großmann ,[*] and Erich Runge 
*Institute of Physics and Institute of Micro- and Nanotechnologies, Technische Universität Ilmenau, 98693 Ilmenau, Germany*

Predicting spectra and related properties such as the dielectric function of crystalline materials based on machine learning has a huge, hitherto unexplored, technological potential. For this reason, we create an *ab initio* database of 9915 dielectric tensors of semiconductors and insulators calculated in the independent-particle approximation (IPA). In addition, we present the OPTIMATE family of machine learning models, a series of graph attention neural networks (GAT) trained to predict the dielectric function and refractive index. OPTIMATE yields accurate prediction of spectra of semiconductors using only their crystal structure. Smooth, artifact-free curves are obtained without these properties being enforced by penalties.

*Introduction.* In recent years, modern machine learning (ML) techniques have been successfully applied to many problems in materials science, ranging from the prediction of various macroscopic properties, such as the formation energy [1–5], the band gap [6–14], the static dielectric constant [15,16], the critical temperature of superconductors [17–19], and many others too numerous to mention here, to the prediction of novel, possibly stable materials [20–25]. Despite the highlighted rapid progress of ML in materials science, the prediction of optical spectra for crystalline materials remains remarkably unexplored, while offering the possibility of finding novel or tailored materials for photovoltaic systems [26], photocatalytic water splitting applications [27–31], epsilon-near-zero materials [32–34], optical sensors, light-emitting devices, and other optical applications. The rarity of ML efforts on frequency-dependent optical properties is due to two major obstacles, which at the same time represent rich research opportunities.

(i) The prediction of spectra is largely unexplored and challenging: For example, the networks of the present work are able to predict the real and imaginary part of the dielectric function of semiconductors calculated within the independent-particle approximation (IPA) for an arbitrary number of $N_\omega = 2001$ frequencies. Depending on the reader's point of view, it may be considered either obvious or surprising that smooth curves are obtained which also approximately fulfill the Kramers-Kronig relations known from complex analysis by learning formally independent values for such a large number of frequencies. Furthermore, it turns out that small numerical artifacts resulting from the finite numerical resolution of the underlying calculations are discarded in the predicted spectra because they differ between the training samples. These observations suggest considerable internal encoding [35] of what might be called the "physically important features of spectra" if they were better understood.

(ii) The other major impediment is the unavailability of datasets and the difficulty to generate sufficiently large and robust datasets from experiments or theory, which is a necessary prerequisite for any ML method. To our knowledge, the largest database of calculated optical spectra to date consists of the absorption spectra of 1116 compounds as part of the Materials Project [36] and the C2DB [37], which covers about 3500 two-dimensional compounds. For molecular compounds, optical and x-ray spectra are more widely available [38] and have led to encouraging first results regarding the application of ML [39–42].

However, crystals are a very different subset of the material space and require very different modeling from the perspectives of theoretical physics and ML. In order to gain insight into whether it is possible to predict frequency-dependent optical spectra of crystalline solids with state-of-the-art deep learning (DL) models, we use *ab initio* calculations to create a database of the frequency-dependent dielectric tensors for 9915 crystalline semiconductors and insulators, from which all other linear optical properties can be calculated.

In this letter, we focus on gapped materials because of the additional complications that arise when calculating the optical spectra of metals. Mainly, metals require extremely dense k-point sampling to accurately resolve intraband and interband transitions around the Fermi energy, requiring either significantly more computational resources or the use of specialized codes [43].

Leveraging our database, we convert all crystal structures to graphs and then train a GAT on a subset of it and evaluate its performance on a different test subset. The trained family of networks, which we name OPTIMATE (short for Optical Materials), together with the graph-generation algorithm, allows for the evaluation of optical spectra directly from atomic positions in a fraction of the time of expensive *ab initio*

---

calculations, while still capturing the key physical features and delivering satisfying quantitative accuracy.

*Database.* To create the *ab initio* database necessary for the application of ML techniques, we perform high-throughput calculations of the dielectric function for 9915 compounds. The structures are taken from the Alexandria database [24,44] of theoretically stable crystals, filtered by the following criteria: (i) only main-group elements from the first five rows of the periodic table; (ii) less than 13 atoms in the unit cell; (iii) energy distance from the convex hull per atom less than 50 meV times the number of atoms in the unit cell (which is less restrictive for larger cells, for which numerical uncertainties are generally larger); (iv) (indirect) band gap greater than 500 meV. The somewhat-arbitrary maximum of 13 atoms limits the necessary complexity of the graph network, and the limited choice of atomic constituents primarily avoids the issue of strongly correlated electrons [45], which often invalidates simple IPA pictures. Figure S1 of the Supplemental Material (SM) [46] shows the prevalence of individual elements in the training and test set. The structures are then reduced to their primitive standard structure according to the algorithm used in Refs. [47,48]. Density functional theory (DFT) calculations were performed with the plane-wave code QUANTUM ESPRESSO [49,50] using PBE [51] as exchange-correlation functional and optimized norm-conserving Vanderbilt pseudopotentials from the SG15 library (version 1.2) [52]. Note that our band gaps may differ slightly from those in the Alexandria database, since the Alexandria database was created with the VASP code [53,54], using different pseudopotentials. Of the 10 189 compounds in the Alexandria database which fulfill the criteria listed above, 274 were identified to be metals when using QUANTUM ESPRESSO and SG15 pseudopotentials and were therefore excluded.

First, the DFT calculations were converged using the protocol described previously in Refs. [47,48]. Then the k-point grid was shifted off symmetry and the $xx$ component of the dielectric tensor was calculated in the IPA on an energy grid consisting of $N_\omega = 2001$ equally spaced points in the energy range 0–20 eV as implemented in YAMBO [55,56]. At this stage, all spectra were calculated with a broadening parameter $\eta = 100$ meV [56]. In the next step, the density $n_k$ of the shifted k-point grid density was iteratively increased until the similarity between the $xx$ component of Im($\varepsilon$) was sufficiently high. To quantify the similarity between frequency-dependent properties, we use the similarity coefficient SC [57], which is defined as follows:

$$\text{SC}[\tilde{Y}(.); Y(.)] = 1 - \frac{\int |Y(\omega) - \tilde{Y}(\omega)| \, d\omega}{\int |Y(\omega)| \, d\omega}, \quad (1)$$

where $Y(\omega)$, $\tilde{Y}(\omega)$ are the frequency-dependent properties to be compared. For the convergence with respect to $n_k$, we used a SC[Im($\varepsilon(\omega; n_k + \delta n_k)$); Im($\varepsilon(\omega; n_k)$)] threshold of 0.9 (*cf.* Ref. [57], where a smaller threshold of 0.75 is used). Using the converged $n_k$ value, final IPA calculations are performed with two broadening values, $\eta = 100$ meV and 300 meV, to investigate the effect of different broadening values on the performance of the ML models. Additionally, if required by the (lack of) symmetry, the $yy$ and $zz$ components of the dielectric tensor are also computed for both values of $\eta$.

Before using the database for ML, we removed a few outliers with direct gaps above 10 eV, mainly noble gases. Materials with an indirect gap below 500 meV are also removed, as these may be metals where a band crossing may have been missed. After filtering, the final database used for the subsequent ML consists of 9748 different materials.

*Network architecture and training.* With the database now available, we, in the framework of DL, attempt to predict the spectra of new materials simply from trends contained in the dataset, in a fraction of the time that an *ab initio* calculation would take. For this purpose, we apply GATs [58], a further development of graph neural networks which have recently achieved dramatic successes in the field of materials science [59]. Both belong to the class of DL algorithms, as they consist of multiple stacked neural network layers. We convert the crystal structures into multigraphs [8] by creating a node for each atom in the unit cell and creating an edge between nodes if the distance between corresponding atoms is less than 5 Å, taking into account periodic boundary conditions. As initial node embeddings, we use a one-hot encoding of the atom's group and row in the periodic table. As edge embeddings we use the Gaussian expanded bond distance, i.e., we evaluate for distances $r$ between 0 Å and 5 Å using a step size of $\delta = 0.1$ Å a Gaussian distribution with standard deviation $\sigma^2 = 0.05$ Å and mean $\mu = r_{ij}$, where $r_{ij}$ is the distance between the corresponding atoms. $\sigma$ and $\delta$ can be considered as hyperparameters that we did not optimize in this letter. Following the results of Ref. [60], which found only a small improvement in the density of states (DOS) prediction error by using more advanced representations, we simply represent the frequency-dependent optical spectra as an $N_\omega$-dimensional vector.

The architecture of OPTIMATE models is as follows (see Fig. 1 for a sketch of the architecture): First, the node embeddings are passed through a multilayer perceptron (MLP) to learn the embedding and expand the parameter space. Then, the graph attention operator from Ref. [61] is applied in one to three message-passing layers to obtain a high-dimensional node embedding, where the number of message-passing layers used is a hyperparameter. These embeddings are then pooled according to the following operation inspired by the graph attention operator:

$$x_{\text{Graph}} = \sum_i \frac{\exp(\text{ReLU}(\theta x_i + b))}{\sum_j \exp(\text{ReLU}(\theta x_j + b))} x_i, \quad (2)$$

where $x_{\text{Graph}}$ is the final learned representation of the graph, $x_i$ is the learned representation of the $i$th node of the graph, $\theta$ is a learned matrix, $b$ is a learned bias vector, and all operations except for the matrix-vector multiplication are carried out elementwise. The vector $x_{\text{Graph}}$ is then finally passed through a MLP mapping to the $N_\omega$-dimensional output vector. We use rectified linear units (ReLU) as the activation function in all cases.

As noted in Ref. [25], using a simple train-test split on a dataset can result in compounds with the same chemical composition but (sometimes only slightly) different crystal structures being present in both the training and test sets, leading to improved error measures, but not to improved generalization for chemical formulas not present in either set. Therefore, we index all unique compositions in the dataset and randomly select 80% of the compositions for training and 10% each for validation and testing. For example, if the composition $Al_2O_3$ were among those selected for training, all polymorphs of $Al_2O_3$ would be included in the training set and thus excluded from the test set. Detailed information on the loss function, training process, and hyperparameter optimization can be found in the SM [46].

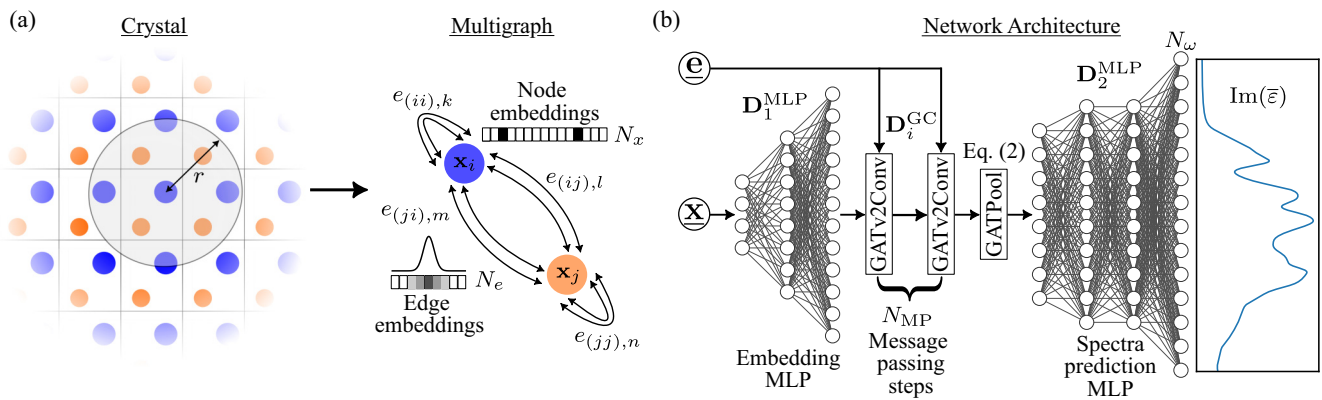

FIG. 1. (a) Schematic illustration of the conversion of a 2D toy crystal into its multigraph representation, chosen for clearer visualization. Connections between atoms are determined by all periodic neighbors within a fixed cutoff radius $r$. The nodes **x** encode atom types by periodic table groups and periods using one-hot encoding, while bond distances are captured by Gaussian expanded-edge embeddings **e**. The approach for the 3D case is analogous. (b) Architecture of the OPTIMATE models, highlighting the integration of encoded atomic and bond information. The dimension of the MLP layers ($D_1^{MLP}$, $D_2^{MLP}$), the number of message passing steps ($N_{MP}$), and the parameters of the $i$th GATv2Conv layers ($D_i^{GC}$) [61] are all architecture parameters that were optimized during the hyperparameter optimization described in the SM [46]. The initial dimension of the node embeddings and the dimension of the edge embeddings are fixed ($N_x = 13$, $N_e = 51$).

*Results.* In this letter, we focus on learning the average of the dielectric function over the diagonal elements of the dielectric tensor, i.e., $\bar{\varepsilon} = \text{Tr}(\boldsymbol{\varepsilon})/3$ and the resulting averaged refractive index defined as $\bar{n} = \sqrt{\bar{\varepsilon}}$. For a total of four OPTIMATE models and for both broadening values, $\eta = 100$ meV and 300 meV, we evaluate $\text{Im}(\bar{\varepsilon}_\eta)$ and $\text{Re}(\bar{n}_\eta)$ as described [we note that OPTIMATE models perform similarly well for Re($\bar{\varepsilon}_\eta$) and Im($\bar{n}_\eta$)]. In Fig. 2, we show a comparison between the *ab initio* spectra and the output of the trained models for the 20%, 40%, 60%, and 80% quantile $Q$ of the test set, as measured by the similarity coefficient SC[DL; IPA]. In Figs. S2–S5 of the SM [46], we show the same comparisons for a larger, randomly selected subset of the test set. Overall, an extraordinary qualitative and a very good quantitative agreement is found; the DFT spectra can be predicted very accurately. First of all, the OPTIMATE shows no aphysical artifacts such as discontinuities, significant noise, or extreme values, even though continuity or smoothness, etc. of the predicted spectra were neither enforced nor promoted during the learning process by, e.g., additional regularizations. A final transposed convolution layer, which would remove such artifacts if they appeared, did not improve the results and does not seem necessary for learning spectral properties.

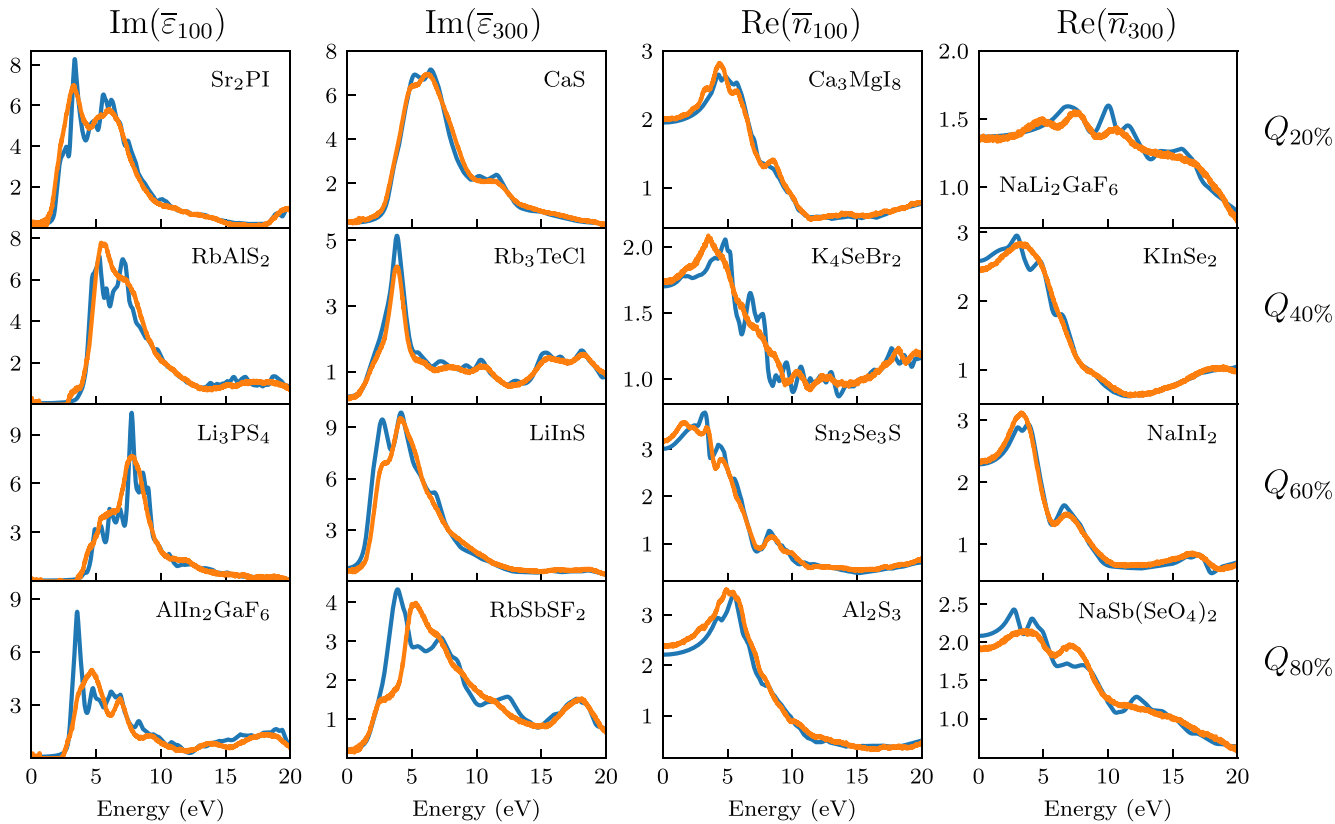

FIG. 2. Comparison between *ab initio* and DL-calculated optical properties on the test set. Each column corresponds to the property shown above. Each row shows the property for a material at that quantile $Q$ of the SC[DL; IPA]. Blue lines show the *ab initio* spectra, orange lines the output of the network. The composition of the material is shown in each cell.

In general, the structure and scale of the optical spectra obtained by OPTIMATE are very similar to those from *ab initio* calculations, especially for compounds with rather simple spectra consisting of only a few strong peaks. In most cases where the intensity of a peak is not predicted well, the position of the peak is still approximately correct. As expected, the OPTIMATE model trained on the data calculated with $\eta = 300$ meV broadening shows significantly less structure than the model trained on the data calculated with an $\eta = 100$ meV broadening. Generally, the output of the models shows less structure than the *ab initio* spectra. This is especially noteworthy for the case of $\eta = 100$ meV broadening, where the spectra of various materials still appear to have significant k-point *ripples* due to insufficient k-point sampling in the *ab initio* calculation, which are often difficult to distinguish from actual transitions (see, e.g., Ref. [62] regarding artifacts due to insufficient k-point sampling). Since k-point *ripples* are essentially an aphysical and hard-to-predict artifact, we interpret this observation as a highly welcome consequence of the fact that the mapping from the "crystal structure to the contribution of k-point *ripples*" is much harder to learn than the physical mapping from the "crystal structure to the dielectric function". Similar observations can be hoped for in general for future predictions of different spectra. We note in passing that the challenges of the model in describing highly structured signals can also be observed in the literature on DOS learning [63]. Remarkably, OPTIMATE is also able to predict the height of the refractive index curve, i.e., the mean refractive index over the frequency range (see Fig. 2). We show several quantitative error measures, i.e., the mean square error (MSE), the mean absolute error (MAE), the mean average percentage error (MAPE), and SC evaluated on the test set for the four OPTIMATE models trained in Table I. A recent publication by Carriço *et al.* [64] predicting the static refractive index, i.e., Re($\bar{n}(\omega = 0)$), using a similar architecture on a different *ab initio* database, obtained a MAPE on the test set of about 4.5–6 %, similar to our results.

Finally, to gain some insight into the difficulty of the task, we evaluate the SC obtained by taking the frequency-resolved mean of the training set as a prediction for the Im($\bar{\varepsilon}_{100}$) and the Re($\bar{n}_{100}$) test sets. This leads to a significantly worse mean (median) similarity coefficient of 0.31 (0.39) and 0.76 (0.78), respectively. Figure 3 shows the distribution of similarity coefficients between this baseline and the OPTIMATE model for the properties calculated with $\eta = 100$ meV (see Fig. S6 for the corresponding $\eta = 300$ meV histograms). A dramatically better prediction of the model compared to the mean is evident. The accuracy of the prediction of Im($\bar{\varepsilon}_{100}$) and Re($\bar{n}_{100}$), measured in terms of the SC, is overwhelmingly between 0.7 and 0.95 and between 0.9 and 1.0, respectively.

TABLE I. Performance of the OPTIMATE models for various tasks. The first value indicates the mean, the value in parentheses indicates the median; both evaluated over the test set. Since $\bar{\varepsilon}$ is (close to) zero below the band gap, the MAPE is meaningless and therefore not given.

|      | Im($\bar{\varepsilon}_{100}$) | Im($\bar{\varepsilon}_{300}$) | Re($\bar{n}_{100}$) | Re($\bar{n}_{300}$) |
|------|---------------|---------------|-------------|-------------|
| MSE  | 0.90 (0.30)   | 0.44 (0.11)   | 0.03 (0.01) | 0.02 (0.01) |
| MAE  | 0.38 (0.33)   | 0.26 (0.20)   | 0.10 (0.09) | 0.08 (0.07) |
| MAPE |               |               | 7.8% (6.7%) | 5.8% (5.0%) |
| SC   | 0.78 (0.81)   | 0.86 (0.88)   | 0.93 (0.94) | 0.94 (0.95) |

*Conclusions.* The results presented in this letter show that DL models commonly used in materials science are able to learn the spectral properties of semiconductors and insulators without significant adaptations to the model architecture and without the need for a much larger database, as might be feared when learning a high-dimensional vectorial quantity instead of a scalar. Significant features are well reproduced. In the case of the refractive index, our frequency-dependent model prediction achieves a quantitative accuracy similar to one recently obtained for the (scalar) static refractive index [64]. As is often the case with machine learning, it is instructive to examine the outliers: the presented OPTIMATE models perform most poorly in predicting the properties of solid fluorine, likely because no other compound contained neutral F atoms. Similarly, planar BN presented challenges. Thus, our research on spectra confirms a motif often observed in ML-based computational materials science and physics in general: The cases where predictions fail deserve a closer look. They may exhibit special physical or chemical properties.

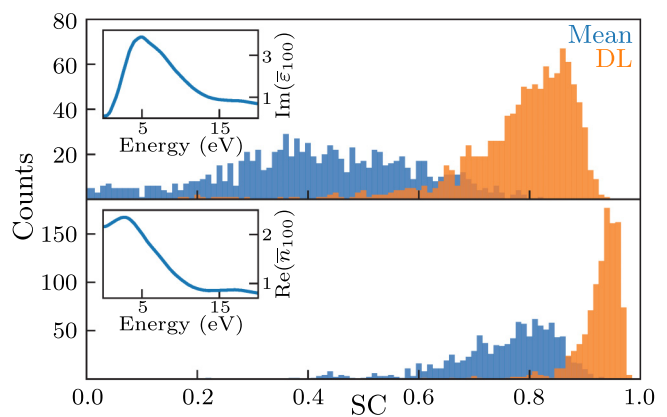

FIG. 3. Histogram of SC[DL; IPA] (orange) and SC[Mean; IPA] (blue) for Im($\bar{\varepsilon}_{100}$) (top) and Re($\bar{n}_{100}$) (bottom) evaluated over the test set, where 'Mean' stands for the frequency-resolved mean of the training set. We note that SC can be negative. The insets show the frequency-resolved mean of Im($\bar{\varepsilon}_{100}$) and Re($\bar{n}_{100}$) over the training set.

The most obvious way to improve the next generation of OPTIMATE models is to increase the size of the dataset used for training, both by including additional elements such as the transition metals or lanthanides (taking into account the challenges of treating them with *ab initio* methods) to allow the network to generalize to more compounds, and by including more compounds containing the elements already included to improve performance. Another promising direction is that instead of using separate models to learn the real and imaginary parts of $\varepsilon$ or $n$, one could design a model that learns both at the same time, since both parts are connected by the Kramers-Kronig relations. Following the principles of physics-informed neural networks, one could incorporate the Kramers-Kronig relations into the loss functions [63,65].

One aspect that should be mentioned is the limitation of the underlying *ab initio* data. We neglect local field effects by using the IPA instead of the random phase approximation (RPA) and completely miss excitonic effects, which can be accounted for by the computationally much more demanding Bethe-Salpeter equation (BSE) [66]. Furthermore, the use of Kohn-Sham DFT eigenvalues leads to a general underestimation of band gaps and thus of the absorption edge, which can be corrected by a more expensive GW calculation [66]. An obvious, but probably suboptimal, next step is to test whether a similar-model architecture can also be trained and applied to data obtained from levels of theory which are closer to experiment but also much more expensive, such as the RPA or GW-BSE, or to high-throughput experimental data [67] of sufficient quality. A more promising approach to circumventing the increased cost of the methods described above would be to test whether transfer learning for spectral properties—building on the presented model—works similarly well as for scalar properties [68], thereby reducing the number of high-quality data points needed. We see no particular reason why this should not be the case.

In summary, we (i) present an *ab initio* database of frequency-dependent optical spectra of crystalline semiconductors and insulators that (ii) allowed us to demonstrate the applicability of current DL techniques to high-dimensional vectorial material properties such as the dielectric function and the refractive index.

The code supporting this publication is publicly available at [69] (IPA calculation workflow) and [70] (OPTIMATE models and analysis). The database of optical spectra for both levels of broadening are available at [71].

*Acknowledgments.* We thank the staff of the Compute Center of the Technische Universität Ilmenau and especially Mr. Henning Schwanbeck for providing an excellent research environment. This work is supported by the Deutsche Forschungsgemeinschaft DFG (Project No. 537033066).